\documentclass{eas}
\usepackage{graphicx}
\begin{document}

%%-----------------------------
%%      the top matter
%%-----------------------------
\title{Instabilities of rotating compact stars: a brief overview} 
\author{Lo\"\i c Villain}\address{CAMK, Bartycka 18, 00-716 Warszawa
(POLAND) and LUTH, Observatoire de Paris - Meudon, 5, place Jules
Janssen, 92195 Meudon (France)}
\begin{abstract}
Direct observations of gravitational waves will open in the near
future new windows on the Universe. Among the expected sources,
instabilities of rotating compact astrophysical objects are waited to
be detected with some impatience as this will sign the birth of
``gravitational waves asteroseismology'', a crucial way to improve our
knowledge of matter equation of state in conditions that cannot be
reproduced in a lab. However, the theoretical work needed to really
get informations from to-be-detected signals is still quite large, numerical
simulations having become a necessary key ingredient. This article
tries to provide a short overview of the main physical topics involved
in this field (general relativity, gravitational waves, instabilities
of rotating fluids, {\it etc.}), concluding with a brief description
of the work that was done in Paris-Meudon Observatory by Silvano
Bonazzola and collaborators.
\end{abstract}
\maketitle
%%-----------------------------
%%      your text
%%-----------------------------
\section{Introduction}

Among the greatest discoveries made in astrophysics during the last
decades, quasars, active galactic nuclei, gamma ray bursts and pulsars are good illustrations
of the fact that general relativity (GR) is getting a growing
importance in explaining physical mechanisms behind high energy
astronomical observations. More precisely, those four classes of
phenomena involve so-called ``compact objects'', whose main properties
can only be understood by taking into account Einstein's theory of
gravitation. With numerous high energy
satellites (XMM-Newton, Integral, Chandra, {\it etc.}) now fully operative, eagerness to
apprehend inner structure and dynamics of these objects or of matter
surrounding them (accretion disks) has been an incitement to deepen
the study of general relativistic hydrodynamics. However, another
reason why the dynamics of fluids in strong gravitational
fields\footnote{{\it i.e.} curved spacetimes whose curvature is
possibly created by the fluid itself.} has naturally become one of the
hottest topics in theoretical astrophysics is that some gravitational
wave (GW) detectors (VIRGO, LIGO, {\it etc.}) are entering into the
game. As this article will illustrate, material compact objects are
indeed very good candidate sources of GWs, the study of their
hydrodynamics being a key to understand how the emission takes
place.\\

 The present text is the written transcription of a talk I had the
pleasure to give at Carg\`ese (Corsica, France) in May 2005. This was
during a School on ``Astrophysical fluid dynamics'' organized by
B\'erang\`ere Dubrulle and Michel Rieutord in honour of Jean-Paul Zahn
and Silvano Bonazzola. The latter has been working in Paris-Meudon
Observatory since 1972, his main topics of interest being relativistic
astrophysics, numerical relativity and gravitational waves. In those
fields, Silvano is mainly known for having initiated the use of
``spectral methods'', making the Meudon group known for the high level
of precision reached in its numerical works (see for instance
Bonazzola {\it et al.} \cite{bona99}). In fact, most of the people
working in that group have had the preparation of their PhD thesis
supervised by him\footnote{for instance E. Gourgoulhon, whose
contribution in this Volume gives a nice introduction to relativistic
hydrodynamics using Cartan external calculus, as was developed by
B. Carter.}, as it was my case from 1999 to 2002. The subject of my
thesis was the study of some gravitational-wave driven instabilities
of rotating neutron stars (NSs), the so-called r-modes, and I was
kindly invited by the organizers of the Carg\`ese school to give a
lecture on that issue. However, this subject and my common work with
Silvano are profitably replaced in the more general context of
instabilities of rotating compact stars, a field in which Silvano had
already been involved previously with some other collaborators from
Meudon (see Section \ref{s:sil}). Hence, in the following, I do not pretend
to be exhaustive and precise, but aim at giving a very short and general
survey of some chosen studies/problems related to the question of
hydrodynamical instabilities of rotating compact stars, a subject in
which the emission of gravitational waves is an essential
phenomenon. Yet, as the school addressed not only to astrophysicists
but also to people working in hydrodynamics without any link to
astrophysics, brief introductions on compact stars and gravitational
waves begin this article, before it deals with oscillations and
instabilities of rotating compact stars, the astrophysical relevance
of the described mechanisms being discussed in the Conclusion.

\section{Compact stars} \label{sec:cp}

Formation, structure and evolution of compact stars are complex
subjects described in detail within various monographs and
review articles such as Shapiro \& Teukolsky (\cite{st83}), Bethe (\cite{bet90}), Pons {\it
et al.} (\cite{prplm99}) and Weber (\cite{w99}). Here, we shall only
give a very brief summary, mainly explaining what is a ``compact
star'' and what are its typical features.\\

 To start with, it is maybe worth reminding that the more massive a
star is at birth, the faster it evolves, producing heavier and heavier
nuclei by successively making fusion of lighter ones. The most stable
element being Fe$^{56}$, no star can use it to produce thermal energy
to counterbalance the gravitational attraction. As a consequence, when
a star is sufficiently massive to produce Fe$^{56}$ nuclei, the latter
begin to accumulate in its core. Being sterile, this iron core owns
its survival mainly to the degeneracy of the electrons that makes the
matter neutral. However, as was shown by Chandrasekhar, a
self-gravitating quantum gas of fermions admits a maximal mass called
the ``Chandrasekhar mass'' (around $1.5$ Solar masses, the exact value
depending on the lepton fraction). If the gas has a mass larger than
this threshold value, the central fermions are so energetic that they
become relativistic, which lowers the compression modulus. This
generates an instability of the iron core, in such a way that when its
mass reaches the Chandrasekhar value, the inner part of the star
suddenly collapses\footnote{Notice that the collapse also results from
the disappearing of some energy due to electron captures and
photodesintegration of iron nuclei taking place at high density.}. Due
to the repulsive nature of the strong interaction at small distances,
the fall of the matter ends with a bounce off of the inner part when
the density reaches values around twice the atomic nuclear density,
also called the ``saturation density''
($n_s\,\sim\,2.5\,\times\,10^{14}\,\textrm{g cm}^{-3}$). Since the outer part is
still falling, a shock-wave is generated that was first thought to be
responsible for the ejection of the external layers with an intense
electromagnetic emission leading to a type Ib, Ic or II
supernova. However, it is now known that the way toward successful
supernova is more complicate, involving also for instance neutrinos
and convection, the full mechanism being still unperfectly
understood. Yet, the supernova being successful or not, the central
remnant left, a compact warm plasma mainly composed of neutrons,
protons, electrons and neutrinos called a ``proto-neutron star'' very
fast begins to contract, while losing energy and neutrinos. Depending
on the amount of matter falling back on it and on the detail of the
dynamics, the proto-neutron stars gives birth to a black hole (BH), a
cold neutron star or a cold strange star (SS), the latter being formed
if the baryonic matter undergoes a phase transition to a quark-gluon
plasma.\\

These classes of astrophysical objects are called ``compact'' due to
the fact that the ratio between their Schwarzschild
radius\footnote{which are proportional to their masses:
$R_s\,\equiv\,2\,M\,G/c^2$} and their radius, called ``compactness
parameter'', is much larger than for usual objects. For instance, a
typical NS has a mass similar to that of the Sun
(\mbox{$M_\odot\,\sim\,2.\,10^{30}$ kg}), but a radius which is $10^5$
smaller (around $10$ km for a NS and $10^6$ km for the Sun). Its
compactness is thus much larger, being around $0.2$ instead of
$10^{-5}$ for the Sun. It is in fact easy to see that the compactness
of any object is by definition smaller than $1$, reaching values close
to it only for NSs (or SSs), and being equal to it only for
BHs. Incidently, it means that the compactness of a star is a
measurement of ``how relativistic it is''. The closest it is to $1$,
the more general relativity is needed to properly describe the star,
its behaviour and what happens around it. In the following, we shall
deal only with the hydrodynamics inside compact stars, {\it i.e.}, we
shall not consider ``pure geometrical'' black holes, in which, in the
spherically symmetric case, all the matter is led to a central
singularity. This restriction has two main consequences on the
hypothesis that can be made in the description of our fluid ball:
\begin{description}
\item[-] it is compact, composed of degenerate nucleons or quarks, gravitation
being mainly counterbalanced by the strong interaction/Quantum ChromoDynamics. Due to the
saturation property of the latter, it can be shown that taking a
constant density profile at null temperature is a reasonable approximation;
\item[-] (almost-)conservation of angular momentum during the collapse
can lead to very high angular velocities, the star possibly reaching
the Kepler angular velocity for which the gravitational force at the
equator is hardly equal to the centrifugal force, meaning that any
additional acceleration would imply a loss of matter.
\end{description}

  As we shall see now, these two characteristics of compact stars make
  them favourable for the emission of gravitational waves.

\section{General relativity and gravitational waves}

Introductions to the physics of gravitational waves and of their
astrophysical sources can be found in the famous monograph {\it Gravitation}
by Misner, Thorne \& Wheeler (\cite{mtw73}), or in the proceedings
Deruelle \& Piran (\cite{der83}), Marck \& Lasota (\cite{ml97}) and
Borane {\it et al.} (\cite{bal00}). Here again, we shall only give a
brief and general summary.\\

\subsection{Stepping stones in GW history}

As soon as 1916, Einstein understood that his new theory of
gravitation, general relativity, could admit solutions describing
ondulatory perturbations of the gravitational field that propagate at
the maximal allowed velocity, $c$. This was the first mathematical
description of ``gravitational waves'', even if as early as 1907,
Poincar\'e had already mentioned the probable existence of some
``ondes gravifiques'' in any relativistic theory of gravitation. In
1918, Einstein came with a formula giving the gravitational emissivity
of a relativistic fluid ball with weak gravitational field and slow
internal motions. As this ``quadrupole formula'' and the solution
describing gravitational waves had been reached in a given system of
coordinates and after linearizing the equations of general relativity,
the question arose of the physical relevance of gravitational
waves. Since GR tells us that all systems of coordinates are
equivalent and none of them directly has a physical meaning, it could
be thought that those waves were just ``coordinate waves'' without any
physical content. That issue started to find an answer only when
Pirani (\cite{p56}) asked the physical question ``What would happen to
my GW detector if a GW goes through my lab?'' and gave an answer to it,
proving that this phenomenon would actually be linked with an energy
deposit. However, the symmetric question (``what does happen to an
object emitting GWs?'') was harder to deal with, and it is only in
the early eighties that it was closed with the demonstration that a
gravitational system emitting GWs does really lose energy until its
total energy reaches a finite positive value. This was done through
various works starting with (e.g.) Bonnor (\cite{bon59}), Bondi {\it
et al.} (\cite{b62}) and ending with (e.g.) Schoen \& Yau
(\cite{s79}), Ludvigsen \& Vickers (\cite{lv81}), Witten (\cite{w81})
and Schoen \& Yau (\cite{s82}). Nevertheless, even before the
existence of GWs inside Einstein theory was theoretically proved,
people had started to experimentally look for them, following a
pioneer called Weber.\\

\begin{figure}[ht]
   \centering
   \includegraphics[height=6cm]{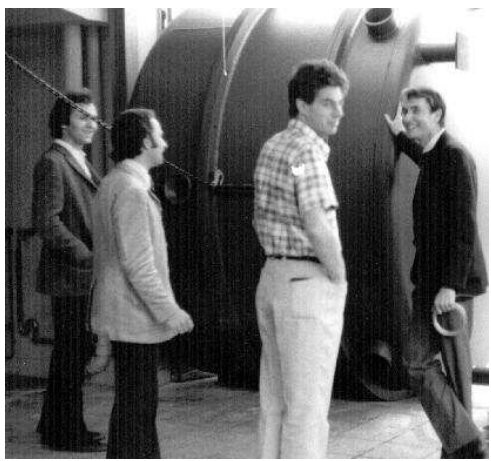}
      \caption{From left to right, Jean Thierry-Mieg, Georges Herpe,
      Silvano Bonazzola and Michel Chevreton in front of the Meudon GW
      detector.}
      \label{md.fig}
\end{figure}

The story of GW detection has indeed its roots in the late sixties
with the building by J. Weber of the first resonant gravitational
detector (a ``bar''), which was some huge and massive bulk whose
eigenmodes should be excited by a coming GW. Shortly after having
begun to take data, Weber effectively announced that he had
successfully detected a signal. This detection was negated later, yet
Weber's work and supposed results had definitely boosted the race for
GW detection, various similar bars being built around the world. As
an example, the Paris-Meudon Observatory experiment can be mentioned,
which was under the responsibility of Silvano and
collaborators (see Fig. \ref{md.fig}) until the end of 1974 when the experiment was given
up.\\

\begin{figure} 
\centerline{\includegraphics[height=6cm]{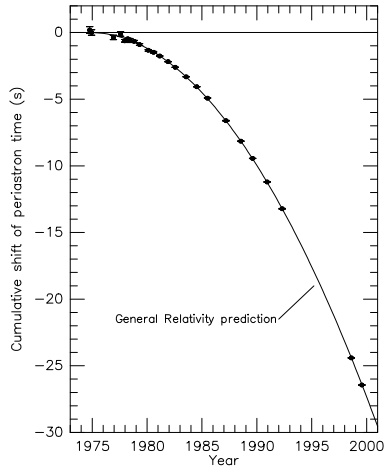}}
\caption{Decreasing of the orbital period of the binary pulsar PSR
B1913+16 measured by the cumulated shift in the time at
periapsis. Straight line curve is the theoretical prediction of GR
[From Lorimer (\cite{l01})].}
\label{f:pb1316}
\end{figure}

 As will be schematically explained in the next section, the first
 evidence of the existence of GWs could only come from astrophysics as
 no terrestrial object can be a relevant source. But quite ironically,
 that first experimental proof came almost exactly at that moment when
 most bar experiments were being left, this proof being based on the
 discovery by Taylor \& Hulse in \cite{th75} of a pulsar in a binary
 system (PSR B1913+16). Indeed, since it is in strong gravitational
 interaction, very precise measurement of the evolution of this pulsar
 could prove that it was behaving exactly as Einstein theory had
 predicted, that is to say with an acceleration of its orbital motion
 resulting from the loss of energy emitted as GWs (see Fig
 \ref{f:pb1316}).\\

 At the beginning of 2006, in spite of the discoveries of some systems
 with much stronger gravitational field, leading to some better tests
 of GR, still no direct detection of GWs has been done. Hence, this is
 obviously one of the main goals of the recent interferometric GW
 detectors (VIRGO, LIGO, LISA, {\it etc.}), whose main advantage with
 respect to bar detectors is to be sensitive to a large range of
 frequencies\footnote{It should be mentioned that there are also
 various bar detectors, much more advanced that Weber's, making with
 the interferometers a whole international network of GW
 detectors.}. Nevertheless, it is worth reminding that even if the
 first direct detection of a GW signal will probably be done by one or
 several of those experiments (now taking data or in commissioning),
 they are above all gravitational telescopes and not detectors to
 prove the existence of GWs.

\subsection{Emission of GWs in GR}

 The only way to really predict the GW signal emitted by a body is to
 solve Einstein equations with, as a source term, the energy-momentum
 tensor of the body and to look at what reaches infinity. However, in
 most of the physical situations, this can not be done analytically
 (mainly due to the strong non-linearities of the theory), and in fact
 even using numerics, this is not such an easy task. Yet, it is not
 needed to exactly solve the equations to be able to find the main
 properties of the oscillatory solutions, as is well-known in the case
 of electromagnetic waves (EWs) in Maxwell theory of electromagnetism
 (EM). Furthermore, as we shall illustrate, looking for the
 differences and similarities between GR and EM turns out to be very
 instructive to understand the basic features of GWs and of their
 emission.\\

 Electromagnetic waves can only be emitted if a given
 distribution of electric charges is evolving in time with a breaking
 of the spherical symmetry. This is related to the vectorial nature of
 the EM field (the photon has a spin 1) and to Gauss theorem, which
 exactly states that if the spherical symmetry is not broken, the
 electric field is constant ($\equiv$ there is no physical scalar part
 in the EM field). Another formulation of that statement is that time
 variations of the ``electric multipoles'' higher or equal to the
 dipole are necessary. In a very similar way, it can be shown that the
 second order tensorial field that describes the gravitational field
 in Einstein theory does not include any vectorial or scalar part that
 would be physical. It means that the graviton is a massless spin 2
 particle, but also that to emit GW, a mass-energy distribution needs
 to evolve in time with breaking of spherical symmetry but not only of
 it: axial symmetry can be preserved if and only if the radial
 distribution is not time invariant. Another equivalent assertion,
 which is by far the most precise, is that while a charge distribution
 needs at least time evolving dipoles to emit EWs, a mass distribution
 needs evolving ``mass multipoles'' at least of the order of the
 quadrupole to emit GWs.\\

\begin{figure} 
\centerline{\includegraphics[height=4.cm]{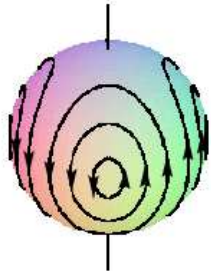}}
\caption{Quadrupolar flow ($l=m=2$)
leading to a current quadrupole and the emission of gravitational
waves. From K. Lockitch PhD thesis}
\label{f:rm}
\end{figure}

 However, we can get a better intuition on GR and GWs by going farther
 in this comparison with EM. Indeed, it is well-known that in Maxwell
 theory, a non-spherical time evolution of a charge distribution is
 not the only way to emit EWs: this can also be done through time
 variations of ``magnetic multipoles'', and not only of electric
 ones. Since magnetic multipoles are equivalent to moving electric
 multipoles, one can expect from this analogy between EM and GR that a
 mass distribution keeping the same shape but having (at least)
 quadrupolar internal motions would also emit GWs through ``current
 multipoles'' or ``gravitomagnetic multipoles''. An easy illustration
 is given by Fig. \ref{f:rm}, which depicts a spherical ball of
 fluid with quadrupolar internal motions\footnote{This can be related with the
 fact that a rotating mass does not generate the same gravitational
 field that a non-rotating one, which is illustrated by the so-called
 ``frame-dragging effect'' that the satellite Gravity Probe B was
 trying to observe close to the Earth. See
 http://einstein.stanford.edu/}. We shall not give here any more
 detail/justification and send the reader to a presentation of the
 ``multipoles expansion'' developed in the framework of
 ``pseudo-Newtonian approaches to GR'' in Thorne (\cite{th80}) or Blanchet
 (\cite{b02}).\\

  Nevertheless, having encountered geometrical criteria necessary for
  getting emission of GWs is not enough. In Nature, if one wants a
  source to emit in a relevant way, its emissivity has also to be
  taken into account. Using again post-Newtonian calculations, it can
  be demonstrated that the gravitational emissivity of a body (with
  mass $M$, typical radius $R$, typical frequency of its internal
  motions $\omega$ and breaking of symmetry $s$) is given by the
  formula
\begin{equation}
\frac{d E}{d t}\,\sim\,\frac{{\cal G_N}}{c^5}\, s^2 \omega^6 M^2 R^4\,,
\end{equation}
where ${\cal G_N}$ is Newton's constant. When one puts in this formula
numbers characteristics of a human size object, {\it i.e.} taking mass,
radius, {\it etc.}, to be around $1$ in I.S. units, the ${\cal
G_N}/c^5$ factor makes that no relevant signal can be emitted since we
have ${\cal G_N}/c^5\sim 3\,\times\,10^{-53}$. In conclusion, no GW can be emitted in
a lab. However, there is a very nice and easy manipulation to do,
which is just to introduce the Schwarzschild radius of the source and a
velocity $v$ typical of its internal motions such that $\omega \sim v /
R$. Putting this velocity in units of the light velocity, one gets
\begin{equation}
\frac{d E}{d t}\,\sim\,\frac{c^5}{{\cal G_N}}\, s^2 \left(\frac{R_s}{R}\right)^2
\left(\frac{v}{c}\right)^6\,.
\end{equation}

 Hence, in ``astrophysically relativistic units'', the $10^{-53}$
 factor has been replaced by its inverse $10^{+53}$, meaning that the
 emission of GWs can be important if the candidate source
\begin{description}
\item[-] is a compact object (high $M/R\,\sim\,R_s/R$),
\item[-] has relativistic internal motions ($v\,\sim\,c$), which are
coherent (to avoid destructive interferences between various
contributions to the time variations of the multipoles).
\end{description}

As a consequence, binary black holes, compact binaries (neutron stars,
strange stars) and isolated compact object oscillating, rotating
(without axial symmetry) and/or accreting are good candidate
sources. However, as far as the last cases are concerned, one can show
that (hydrodynamical) oscillations are damped (at least) by the
emission of GWs, which means that to avoid very short duration
signals, instabilities would be more interesting than just oscillations.

\section{Instabilities of rotating compact stars}

\subsection{Types of instabilities} \label{s:tins}

 The topic of GWs from instabilities of relativistic stars has been
 widely reviewed recently in Andersson (\cite{a03}) and we send the
 reader to this article for more detail. We shall here just summarize
 the main conclusions, starting with the fact that instabilities can
 be classified in two categories: dynamical and secular ones. The
 first ones develop in a timescale that is of the same order as the
 (hydro)dynamical timescale, {\it i.e.} the period of the rotating
 star. While dynamical instabilities are related only to gravitation
 and hydrodynamics, secular instabilities are triggered by some
 dissipative processes acting on much longer
 timescales\footnote{Notice than in GR, gravitation is a dissipative
 process through GWs emission although it is not in Newtonian
 theory.}. In the case of compact objects in GR, the most important
 sources of dissipation are viscosity and emission of gravitational
 radiation. In NSs, the first one mainly results either from
 nucleon-nucleon scattering (dominant process at low temperature) or
 from beta reactions $n\,\leftrightarrow\,p\,+\,e^-\,+\,\nu$ (dominant
 at high temperature), some other mechanisms being also involved, for
 instance related with possible superfluidity of the nucleons,
 interactions between the fluid and the crust, {\it etc.} The main
 difference between dissipation coming from viscosity and dissipation
 linked with emission of GWs is that viscosity does conserve angular
 momentum and reduce vorticity, whereas gravitational radiation takes
 angular momentum away from the star but without changing the
 vorticity. To sum up, a viscosity driven instability leads the star
 from a state of energy $E_0$, angular momentum $L_0$ and
 circulation\footnote{Circulation is the flux integral of vorticity.}
 $\zeta_0$ to a state ($E<E_0,L_0,\zeta<\zeta_0$) and a GW driven
 instability from the same initial state to
 ($E'<E_0,L<L_0,\zeta_0$). As we shall see in the next section, those
 evolutions are made possible by the apparition of some spontaneous
 breaking of symmetry.

\subsection{Equilibrium and instabilities of rotating self-gravitating fluids}

 In Section \ref{sec:cp}, it was noticed that considering compact
 stars as having constant density profiles was not such a bad
 approximation due to the saturation property of strong
 interaction. Yet, since it is the easiest way a star can be modelized
 and since it enables analytical calculations, the uniform density
 model had been used to describe stars much before we had very precise
 ideas about their inner structure (see Chandrasekhar \cite{c69}). As a
 consequence, many results about stability of rotating stars have been
 obtained in the Newtonian framework, which have become classical and
 which are useful to have in mind before looking at the situation of
 relativistic stratified rotating compact stars (see Andersson
 \cite{a03} for a review).\\

 The most important of those results is probably the fact that
 axisymmetric configurations are not always the most natural for
 rotating self-gravitating fluids. Indeed, even if at low angular
 velocity, a uniform density fluid adopts a Mac Laurin spheroid shape,
 when its angular velocity overreaches some value, the spheroid can
 become instable and some stable triaxial ellipsoidal solutions
 appear. Usually, the parameter used to characterize the configuration
 is not the angular velocity, but the ratio between the (rotational)
 kinetic energy and the gravitational energy. This parameter, called
 $\beta$ in the literature, can be shown to be always smaller than
 $0.5$ due to the virial theorem. The relevance of $\beta$ and the
 preference for triaxial configurations at large angular velocities can
 be understood in the origin of the instabilities, which is the
 competition between two forms of energy: rotational and
 gravitational. As an example, take a star with a given angular
 momentum $J$ that is supposed to be conserved. The narrower its mass
 distribution, the smaller its moment of inertia $I$ and the larger
 its rotational energy $T$, since it can be written
 $T=\,J^2/(2\,I)$. Nonetheless, its gravitational energy $K$ is an
 increasing function of the mass distribution size, meaning that $T$
 and $K$ have opposite behaviours. Hence, the larger is the angular
 momentum of a star, the more its energy will be dominated by the
 rotational part and the more stable will be the configurations with a
 larger moment of inertia, implying that triaxial configurations become
 privileged when the angular momentum is increasing. Another way to
 formulate this result is to say that when the value of angular
 momentum is sufficiently high, there is a critical value of $\beta$
 such that for any $\beta$ larger than that critical value, some
 triaxial configuration is a state of lower energy than the
 axisymmetric one. The same line of argumentation can be followed with
 circulation instead of angular momentum, leading to the conclusion
 that there are two types of triaxial configurations relevant for
 increasing angular velocity, both of them being associated with some
 dissipative process driving the fluid to a state of lower
 energy. However, before giving more detail on this, let us notice
 that, using some ``free-energy approach'', Bertin \& Radicati
 (\cite{br76}) and Christodoulou {\it et al.} (\cite{ckst95}) have
 shown that those transitions\footnote{and also a dynamical
 instability which occurs in the case of inviscid fluids at larger
 $\beta$} can be understood as some spontaneous breaking of
 symmetries, with for order parameter one minus the ellipticity of the
 star, {\it i.e.} $1-x$ where $x$ is the ratio between the lengths of
 the two axis orthogonal to the rotation axis.\\

 As reminded in Section \ref{s:tins}, viscosity conserves angular
 momentum while dissipating vorticity/circulation. Hence, an unstable
 axisymmetric star under the influence of viscosity will be led to the
 first class of ellipsoids called Jacobi ellipsoids, which are rigidly
 rotating configurations with same angular momentum as the initial
 configuration but smaller energy and without circulation. As these
 configurations are triaxial and rigidly rotating, they do have time
 varying mass quadrupoles and can emit GWs, while viscosity no longer
 influences them. The second class is called Dedekind ellipsoids and
 is the final state of unstable axisymmetric stars led to instability
 by GWs emission. Those ellipsoids are characterized by their apparent
 immobility (in the inertial frame) and constant vorticity. Their
 triaxial shape is supported by some internal fluxes and differential
 rotation. As a consequence, they do not emit GWs, but can lose energy
 due to viscosity.\\

\begin{figure} 
\centerline{\includegraphics[height=6cm]{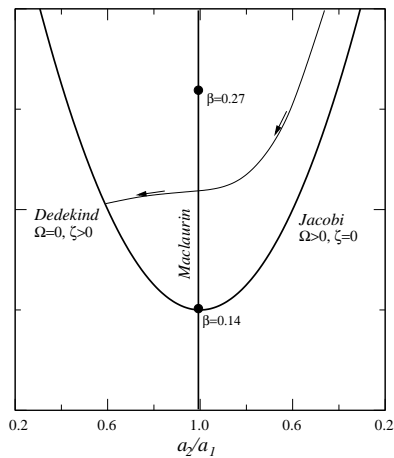}}
\caption{Schematic illustration of the two secular instabilities
discussed here as function of the $\beta$ ratio (see text for
definition) and the ellipticity $a_1/a_2$ (ratio between the axis
orthogonal to the rotation axis). The critical value for both types of
secular instability is around $0.14$ (exactly equal for constant
density Newtonian fluids), while dynamical instability is associated
with a higher critical value $\beta_c\,\sim\,0.27$. As explained in
the text, Jacobi ellipsoids are favoured by viscosity, while
gravitational radiation drives the fluid toward a Dedekind
ellipsoid. The evolution pictured in the figure is an example of a
situation in which the GW driven instability wins. From Andersson
(\cite{a03})}
\label{f:instab}
\end{figure}

 The possible appearance of these two types of triaxial configurations
 is illustrated in Fig. \ref{f:instab} (extracted from Andersson
 \cite{a03} to which the reader is sent for more detail), where has
 been taken into account another important result: the fact that these
 two instabilities admit the same critical value of $\beta$ for
 incompressible Newtonian stars. However, this result should not make
 us forget that the relevance of those instabilities in physical stars
 is with no doubt an issue much more complex than what was briefly
 presented here, at least since compact stars are relativistic and
 compressible. Yet, this conclusion is also supported by the work of
 Lindblom \& Detweiler (\cite{ld77}) who have shown that due to the
 antagonistic roles of viscosity and GWs emission, the two secular
 instabilities of Mac Laurin spheroids discussed here tend to cancel
 each other, the ratio of their strengths being the key to decide
 which of them has the final answer. Before explaining how Silvano and
 collaborators tried to give some elements of answer to that question
 and to the question of the relevance of GW emission by rotating
 compact stars, we shall now try to clarify the link between those
 instabilities and some oscillatory modes of rotating compact stars.

\section{Oscillations and instabilities}

 Since oscillations of self-gravitating fluids have been the subject
 of numerous studies and books, we shall just give here two classical
 references: Unno {\it et al.} (\cite{u79}) and Kokkotas \& Schmidt
 (\cite{ks99}), the second one dealing with relativistic compact
 objects. However, most of the topics discussed in the present article
 are deeper developed in Andersson (\cite{a03}).

\subsection{Effects of rotation on oscillatory modes of stars}

  To illustrate the connection between instabilities and oscillations
of rotating stars, it is worth first reminding the main effects of
rotation on oscillatory modes. As we are dealing with stars without
any other anisotropic physics than rotation (we neglect stress in the
crust, magnetic field, {\it etc.}), it is useful to work with
spherical harmonics decomposition of the functions that describe the
modes. With this approach, a mode of oscillation is characterized by
two ``quantum'' numbers $(l,m)$ such that $m\,\in[-l,l]$, its
frequency being $w_{lm}$. In a spherically symmetric situation ({\it
i.e.} without any rotation), it is well-known that all modes with the
same azimuthal number $m$ do have the same frequency
$w_{lm}\,\equiv\,w_l$. An important physical quantity reached from
this frequency is the ``pattern speed'', which depicts the apparent
deplacement of the wave associated to the mode with respect to the
star. For positive $m$, we get the pattern speed
$\sigma_-\,=\,-\frac{w}{m}\,=\,\frac{d \phi}{d t} < 0$, meaning that
the mode is ``retrograde'', the wave moving in the direction of
decreasing $\phi$. For negative $m$, the pattern speed is
$\sigma_+\,=\,-\frac{w}{m}\,=\,\frac{d \phi}{d t} > 0$, meaning that
those modes are prograde.\\

 When the star is rotating, all the physics is made more complicated,
 from the inner structure calculation (see Stergioulas \cite{st03} for
 a review of rotating stars in relativity) to the behaviour of the
 modes which become coupled. However, even before this coupling, the
 main effects of rotation are the splitting and modification of the
 frequencies, the frequency in the inertial frame ($w_i$) being
 related to the frequency in the frame corotating with the star at
 angular velocity $\Omega$ ($w_r$) by
\begin{equation} \label{eq:mod}
 w_i\,=\,w_r\,-\,m\,\Omega\,+\,C_{lm}(\Omega)\,.
\end{equation}

 In this formula,
 $C_{lm}(\Omega)$ comes from the modification of the inner structure
 and is thus a term of second order in $\Omega$. Due to the
 $m\,\Omega$ term, prograde and retrograde modes are affected
 differently, the sign of their frequencies possibly changing,
 meaning a change of their apparent direction of propagation. As we shall see now,
 this phenomenon is an indicator of instabilities.

\subsection{Oscillations and criteria for finding instabilities}

 Telling if a given equilibrium configuration of a self-gravitating
 fluid is stable or not has always been a complicate problem, the
 negative answer being however easier to prove. Indeed, in this case,
 it is sufficient to find a criteria that can indicate the existence of
 an instability, and not necessarily of all of them. As far as GR and GWs are concerned, an important step
 forward was done during the seventies by Chandrasekhar, Friedman and
 Schutz leading to what is now known as the CFS criteria for
 instability (see for instance Chandrasekhar \cite{c70}, Friedman \&
 Schutz \cite{fs75}, \cite{fs75b}b). What they discovered is that, in GR as
 in the Newtonian and post-Newtonian studies, it was possible to find
 non-axisymmetric instabilities of relativistic rotating stars by
 defining some ``canonical energy'' and looking for the appearance of
 ``neutral modes'' (modes with null frequencies). We shall now
 illustrate this with the modes associated to the two secular
 instabilities presented earlier and with another one discovered at
 the end of the nineties by Andersson (\cite{a98}).\\

\subsubsection{Instabilities and triaxial configurations}

\begin{figure}
   \centering
   \includegraphics[height=6cm]{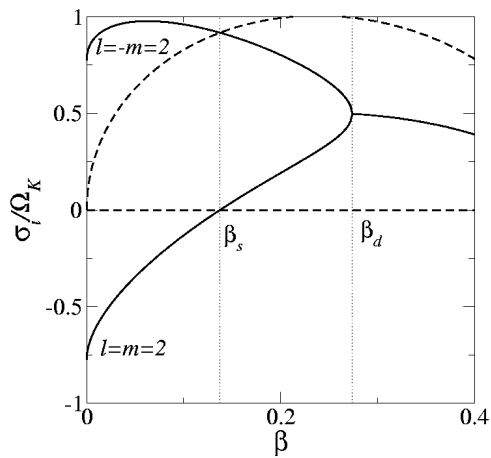}
      \caption{Pattern speeds in the inertial frame of the $l=|m|=2$
      f-modes of a Mac Laurin spheroid as functions of the $\beta$
      parameter. Additionally, in dashed-line is depicted the
      ``vanishing pattern speed in the rotating frame''. It is seen
      that the $m=2$ retrograde mode is driven to instability for
      $\beta=\beta_s$ because its pattern speed vanishes. As explained
      in the text, for this neutral point is obtained in the
      inertial frame, it is linked with a GW driven instability
      leading to Dedekind ellipsoid. However, since the Mac Laurin
      spheroid has a constant density profile and is Newtonian, we
      observe that the prograde mode is driven to instability by
      viscosity for the same value of $\beta$, which is observed by the
      crossing of its pattern speed curve with the (dashed) curve of
      ``vanishing pattern speed in the rotating frame''. Another
      observation that can be made albeit we shall not discuss it here
      is the fact that the appearance of the dynamical instability
      ($\beta=\beta_d$) can be found by the point at which the pattern
      speeds in the inertial frame of the two modes coincide. Finally,
      notice that all frequency are normalized by the Kepler angular
      velocity.}
      \label{f:mcmd}
\end{figure}

 A triaxial shape is associated with $l=|m|=2$ spherical harmonics
 numbers. As a consequence, it should not be a surprise that the
 so-called ``bar-mode'' ($l=|m|=2$ f-mode) is the one that is driven
 to instability by viscosity or by GW emission in the case of
 (respectively) Jacobi or Dedekind ellipsoids appearance. Indeed,
 f-modes are mainly surface waves, and the $l=|m|=2$ modes will thus
 correspond to surface deformations of the same shape as the
 ellipsoids. The only possibly missing point in this intuitive
 argumentation is that for the mode to imply a deformation of the Mac
 Laurin spheroid toward one of those ellipsoids, a synchronization
 between the mode and the ellipsoid to come seems to be needed for
 energy to transfer. In the case of the Dedekind ellipsoid, which has
 a fixed shape in the inertial frame, the mode needs to have a
 vanishing pattern speed in the inertial frame, while in the case of
 Jacobi ellipsoid (rigidly rotating), the mode should have a null
 frequency in the rotating frame. Those conclusions can be shown to be
 in agreement with calculations done using the canonical energy
 formalism developed by Friedman and Schutz, a null frequency meaning
 a change of sign of that energy. The resulting evolution of the
 bar-modes pattern speeds as functions of $\beta$ (or the angular
 velocity) are depicted on Fig. \ref{f:mcmd} (from Andersson
 \cite{a03} in which more detail can be found).\\

\subsubsection{GW driven instabilities and multipoles}

  The CFS criteria very briefly introduced at the beginning of the
  current section has a very crucial implication that we should now
  discuss. Indeed, even if the frequency of any mode is modified by
  rotation, this shift [$C_{lm}$ in Eq.(\ref{eq:mod})] will never
  prevent the existence of a neutral point in the inertial frame. As a
  consequence, for all modes there is a minimal angular of the star
  $\Omega_{lm}$ such that the mode is driven to instability by GW
  emission if the star rotates faster than this. Said in another way:
  all relativistic rotating stars are generically instable. However,
  this conclusion is valid only for inviscid fluids. As already
  explained, viscosity makes the game more complex, and in practice, it
  appears that for increasing $m$
\begin{description}
\item[-] the $\Omega_{lm}$ value decreases, making the instability easier to reach
\item[-] the viscosity timescale decreases, so that viscosity can kill the candidate instability faster
\item[-] the growing timescale for the instability increases, making
it more difficult to observe\footnote{this result comes from
post-Newtonian multipoles calculation}.
\end{description}

\begin{figure}[ht]
   \centering
   \includegraphics[height=6cm]{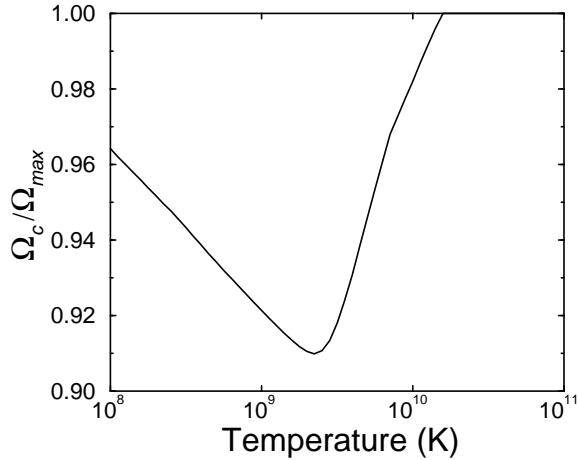}
      \caption{Window of instability for the $l=m=2$ f-mode with
the angular velocity in units of the Kepler angular velocity (maximal
angular velocity at which gravitation can balance centrifugal force at
the equator). Notice that the minimal value of the angular velocity
for instability is $>90\%\,\Omega_K$. From Andersson (\cite{a03})}  \label{f:fmi}
\end{figure}

 Thus, it can be shown that only a few modes are actual candidates for
 GW driven instability, each of them being characterized by a ``window
 of instability'' in the {\it (Temperature, Angular velocity)}
 plane. This is illustrated in Fig. \ref{f:fmi} with the example of
 the $l=m=2$ {\it f}-mode which was quickly recognized as a good
 candidate due to its quite short timescale for the growing of the
 instability. However, as appeared in that figure, the star needs to
 rotate very fast (more than $90\%$ of its Kepler velocity) for the
 instability to grow.\\

\begin{figure}[ht]
   \centering
   \includegraphics[height=6cm]{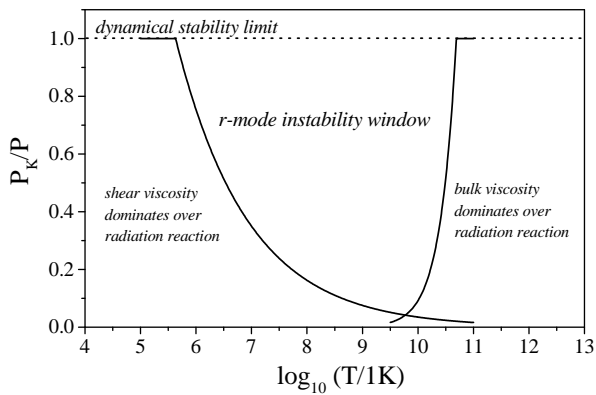}
      \caption{Window of instability for the $l=m=2$ r-mode. From Andersson (\cite{a03})}  \label{f:rmi}
\end{figure}

  The situation slightly changed in 1998 with a discovery made by
  Andersson (\cite{a98}) who found that even if non-axisymmetric modes
  without huge density perturbation fluxes have consequently small
  mass multipoles and can seem not very promising for the emission
  of GWs, they are maybe more so than expected. More precisely, he
  observed that some purely axial inertial modes\footnote{an axial
  mode is a mode whose spherical harmonics decomposition is such that
  for a given $l$ the parity of the mode is $(-1)^{l+1}$. An inertial
  mode is a mode whose restoring force is Coriolis force.} were
  instable whatever the angular velocity of the star: they are always
  prograde in the inertial frame and retrograde in the rotating one,
  emitting GWs through current multipoles. Yet, this was just the
  beginning of the story of r-modes since the question of their
  relevance in actual stars is still open, the importance of viscosity
  strongly depending on the not-so-well-known equation of state and on
  other unknown features of NSs (see Fig. \ref{f:rmi} to compare the
  size of the window of instability of the $l=m=2$ r-mode with that of
  the $l=m=2$ f-mode, which was the best candidate for GWs emission
  before 1998). But in a more general way, it is the whole problem of
  instabilities of rotating compact stars that still relies on
  unknowns linked with microphysics. We shall give now some words on
  trying-to-be realistic studies Silvano performed with various
  collaborators.\\

\subsection{Some works by Silvano and collaborators} \label{s:sil}

All contributions of Silvano to the study of rotating relativistic
stars instabilities that will be discussed now share the use of the
numerical methods he had the great idea to develop in numerical
relativistic astrophysics, the already mentioned spectral methods.

\subsubsection{On viscosity driven instabilities}

Viscosity driven instabilities lead instable Mac Laurin spheroids to
rigidly rotating Jacobi ellipsoids. However, these ellipsoidal shapes
are exact solutions only in Newtonian gravitation and with
constant density profiles. Silvano and his collaborators (J. Frieben
and E. Gourgoulhon) were the first to study the stability of rotating
compact stars with realistic equations of states in general relativity
(Bonazzola {\it et al.} \cite{bon96}, \cite{bon98}). Using a slightly
modified version of a code well-known from people working in numerical
relativity (Bonazzola {\it et al.} \cite{bon93}), they proved that
relativity had a stabilizing effect of self-gravitating rotating
fluids with respect to the viscosity driven instability (in other
words, $\beta_c$ was found larger than in Newtonian theory), while
also testing the precision of the code by looking for the maximal
adiabatic index of a Newtonian polytrope\footnote{It was indeed known
that for a Newtonian polytrope, {\it i.e.} an equation of state for which
pressure is related to density by a power law, if it is too
compressible, the Kepler limit is reached before any viscosity driven
instability can appear.}. To end with viscosity driven instabilities,
it can be mentioned that a step further in the development of the
numerical strategy was covered by several common collaborators of
Silvano by using a multi-domain approach which enable spectral
algorithms to deal with discontinuous functions (Gondek-Rosi\'nska \&
Gourgoulhon \cite{g02}, Gondek-Rosi\'nska {\it et al.} \cite{g03}). In
this way, they were able to get more digits in the calculation of the
standard problems of incompressible Newtonian fluids, but also to
study the case of strange stars for which a surface density
discontinuity must be taken into account.

\subsubsection{On gravitational emission driven instabilities}

The last subject that will be mentioned as an illustration of
Silvano's achievements in the topic of rotating relativistic stars
instabilities is what I had the gladness to work on in collaboration
with him, mainly the instability of r-modes. The approach to that
instability was different than what he did earlier due to the fact
that it does not lead the star to a known state. Thus, the strategy
was to develop an hydrodynamical spectral code to solve the
relativistic Euler equation in realistic NSs. The first version of the
code had a basic description of the microphysics (Villain \& Bonazzola
\cite{v02}) while we later improved it in collaboration with
P. Haensel (Villain {\it et al.} \cite{v05}) to take into account
stratification and frozen composition. It enabled us to make the first
relativistic study of the coupling between gravity modes coming from
stratification and r-modes. However, even if some influence on the
latter was highlighted, as any work that can be done nowadays is much
too imperfect, the question of the relevance of their instability
from the point of view of GW emission is still quite open as we shall
sum up in the conclusion.

\section{Conclusion: Astrophysical relevance of the instabilities}

  As conservation of angular momentum says that compact stars are born
  with quite a fast rotation, all instabilities presented in this
  article (and several more) could develop in newly born NSs, unless
  they are generically forbidden by some phenomenon\footnote{during
  some time, high viscosity resulting from superfluidity of hyperons
  was suggested to play this role, but it seems now that some numbers
  had been overestimated.}. Yet, among the many unknowns concerning
  young compact objects, crucial issues are their profiles of rotation
  and their actual angular velocities at birth\footnote{related to
  these topics also can be mentioned some works based on Silvano's
  development of spectral methods in Meudon: Goussard, Haensel \&
  Zdunik (\cite{g97}), (\cite{g98}) and Villain {\it et al.}
  (\cite{vp04}).}. Indeed, the importance of processes like magnetic
  braking to slow down young compact stars is still under question,
  while instabilities could also be born in proto-neutron stars
  resulting from neutron star binaries coalescences. However, the role
  of differential rotation is not clear either since it allows larger
  values of the $\beta$ parameter, but also possibly creates new
  instabilities appearing at low $\beta$. Consequently, young compact
  stars are still the most natural objects from which instabilities
  sources of GWs can be expected, but as was proposed by Wagoner and
  others, those instabilities could also be influential in low mass
  X-rays binaries (LMXB), systems in which old compact objects
  continuously accrete matter from low mass star. This could be the
  way to a long periodic gravitational signal and from it the
  definitive birth of ``asteroseismology of compact stars'' (For more
  detail on these scenarios and other, see Kokkotas \& Stergioulas
  \cite{ks05}).

\acknowledgements I am grateful to Richard Taillet, Jos\'e Pons and
 Christian D. Ott for various useful comments, to Nils Andersson and
 others for doing nice pictures, but I would also like to thank here
 again B\'erang\`ere Dubrulle and Michel Rieutord for organizing such
 a nice and interesting school in Carg\`ese. Last but not least, I
 should not forget Silvano himself for all past, present and future
 pleasant collaborations and conversations. The writing of those notes
 benefited from the support of the Departamento de F\'\i{}sica
 Aplicada (DFA) of Alicante University.

%%-----------------------------
%%      your bibliography
%%-----------------------------


\begin{thebibliography}{99}

\bibitem[1998]{a98}
Andersson, N., 1998, ApJ, 502, 708

\bibitem[2003]{a03}
Andersson, N., 2003, Class Quantum Grav., 20, R105

\bibitem[1976]{br76}
Bertin, G., \& Radicati, L.A., 1976, ApJ, 206, 815

\bibitem[1990]{bet90}
Bethe, H.A., 1990, Rev. Mod. Phys., 62, 801

\bibitem[2002]{b02}
Blanchet, L., 2002, Gravitational Radiation from Post-Newtonian Sources
and Inspiralling Compact Binaries, Living Reviews in Relativity, 5,
online article: http://relativity.livingreviews.org/Articles/lrr-2002-3/index.html

\bibitem[1993]{bon93}
Bonazzola, S., Gourgoulhon, E., Salgado, M., \& Marck, J.-A., 1993, A\&A, 278, 421

\bibitem[1996]{bon96}
Bonazzola, S., Frieben, J., \& Gourgoulhon, E., 1996, ApJ, 460, 379

\bibitem[1998]{bon98}
Bonazzola, S., Frieben, J., \& Gourgoulhon, E., 1998, A\&A, 331, 280

\bibitem[1962]{b62}
Bondi, H., Van der Burg, M.~G.~J., \& Metzner, A.~W.~K., 1962, Proc. R. Soc. London A, 269, 21

\bibitem[1959]{bon59}
Bonnor, W.~B., 1959, Philos. Trans. R. Soc. London A, 251, 233


\bibitem[1999]{bona99}
Bonazzola, S., Gourgoulhon, E., \& Marck, J.-A, 1999, J. Comput. Appl. Math., 109, 433

\bibitem[2000]{bal00}
Barone, M., \etal, 2000, Experimental Physics of Gravitational Waves, (World Scientific Publishing, Singapour)

\bibitem[1969]{c69}
Chandrasekhar, S., 1969, Ellipsoidal figures of equilibrium, (Yale University Press)

\bibitem[1970]{c70}
Chandrasekhar, S., 1970, PRL, 24, 611


\bibitem[1995]{ckst95}
Christodoulou, D.~M., Kazanas, D., Shlosman, I., \& Tohline, J.~E., 1995, ApJ, 446, 472

\bibitem[1983]{der83}
Deruelle, N., \& Piran, T., 1983, Gravitational
radiation, Proceedings of the Advanced Study Institute, Les Houches,
Haute-Savoie, France, June 2-21, 1982

\bibitem[1975]{fs75}
Friedman, J.~L., \& Schutz, B.~F., 1975, ApJ Lett., 199, L157

\bibitem[1975]{fs75b}
Friedman, J.~L., \& Schutz, B.~F., 1975, ApJ, 200, 204

\bibitem[2002]{g02}
Gondek-Rosi\'nska, D., \& Gourgoulhon, E., 2002, PRD, 66, 044021

\bibitem[2003]{g03}
Gondek-Rosi\'nska, D., Gourgoulhon, E., \& Haensel, P., 2003, A\&A, 412, 777

\bibitem[1997]{g97}
Goussard, J.~O., Haensel, P., \& Zdunik, J.~L., 1997, A\&A, 321, 822

\bibitem[1998]{g98}
Goussard, J.~O., Haensel, P., \& Zdunik, J.~L., 1998, A\&A, 330, 1005

\bibitem[1975]{th75}
Hulse, R.~A., \& Taylor, J.~H., 1975, ApJ Letter, 195, L51

\bibitem[1999]{ks99}
Kokkotas, K. \& Schmidt, B., 1999, Quasi-Normal Modes of Stars and Black Holes, Living Reviews in Relativity, 2,
on line article: http://relativity.livingreviews.org/Articles/lrr-1999-2/

\bibitem[2005]{ks05}
Kokkotas, K.D., \& Stergioulas, N., 2005, Gravitational Waves from Compact Sources, Proceedings of the 5th International Workshop "New Worlds in Astroparticle Physics", Faro, Portugal, 8-10 January 2005 (gr-qc/0506083)
 
\bibitem[1977]{ld77}
Lindblom, L., \& Detweiler, S.L., 1977, ApJ, 211, 565

\bibitem[2001]{l01}
Lorimer, D.~R., 2001, Binary and millisecond pulsars at the new millennium, Living Reviews in Relativity, 4,
online article: http://www.livingreviews.org/Articles/Volume4/2001-5lorimer

\bibitem[1981]{lv81}
Ludvigsen, M., \& Vickers, J., 1981, J. Phys. A 14, L389

\bibitem[1997]{ml97}
Marck, J.-A., \& Lasota, J.-P., 1997, Relativistic Gravitation and Gravitational Radiation, Proceedings of the Advanced Study Institute, Les Houches, Haute-Savoie, France, 

\bibitem[1973]{mtw73}
Misner, C.~W., Thorne, K.~S., \& Wheeler, J.~A., 1973, Gravitation, (W.H. Freeman and Company, San Fransisco)

\bibitem[1956]{p56}
Pirani, F.~A.~E., 1956, Acta Phys. Pol., 15, 389

\bibitem[1999]{prplm99}
Pons, J. A., \etal, 1999, ApJ, 513, 780


\bibitem[1979]{s79}
Schoen, R. \& Yau, S.~T., 1979, Commun. Math. Phys. 65, 45

\bibitem[1982]{s82}
Schoen, R. \& Yau, S.~T., 1982, Physical Review Letters, 48, 369


  \bibitem[1983]{st83}
   Shapiro, S.L., \& Teukolsky S.A., 1983,
Black Holes, White Dwarfs and Neutron Stars,
   (Wiley-Interscience, New-York)

\bibitem[2003]{st03}
Stergioulas, N., 2003, Rotating Stars in Relativity, Living Rev. Relativity, 6,
online article: http://www.livingreviews.org/lrr-2003-3

\bibitem[1980]{th80}
Thorne, K., 1980, Rev. Mod. Phys., 52, 299

\bibitem[1979]{u79}
Unno, W., Osaki, Y., Ando, H., \& Shibahashi, H., 1979, Nonradial oscillations of stars, (Univ. of Tokyo Press, Tokyo)

\bibitem[2002]{v02}
Villain, L. \& Bonazzola, S., 2002, PRD, 66, 123001

\bibitem[2004]{vp04}
Villain, L., Pons, J.~A., Cerd{\'a}-Dur{\'a}n, P., \& Gourgoulhon, E., 2004, A\&A, 418, 283

\bibitem[2005]{v05}
Villain, L., Bonazzola, S., \& Haensel, P., 2005, PRD, 71, 083001

\bibitem[1999]{w99} Weber, F., 1999, Pulsars as Astrophysical
Laboratories for Nuclear and Particle Physics, High Energy Physics,
Cosmology and Gravitation Series, (IOP Publishing, Bristol)

\bibitem[1981]{w81}
Witten, E., 1981, Commun. Math. Phys., 80, 381

\end{thebibliography}
\end{document}